\documentclass{article}
\usepackage{graphicx} 
\usepackage{float}
\usepackage[utf8]{inputenc}
\usepackage[english]{babel}
\usepackage{csquotes}
\usepackage{amsmath}

\usepackage[
backend=biber,
style=alphabetic,
sorting=ynt
]{biblatex}

\title{\textbf{Towards an Ethical and Inclusive Implementation of Artificial Intelligence in Organizations: A Multidimensional Framework}}
\author{Ernesto Giralt Hernández \\ \textit{ernesto.giralt@aquarelleai.com} }

\date{Barcelona, April 2024}

\begin{document}
\maketitle

\begin{abstract}
This article analyzes the impact of artificial intelligence (AI) on contemporary society and the importance of adopting an ethical approach to its development and implementation within organizations. It examines the technocritical perspective of some philosophers and researchers, who warn of the risks of excessive technologization that could undermine human autonomy. However, the article also acknowledges the active role that various actors, such as governments, academics, and civil society, can play in shaping the development of AI aligned with human and social values.

A multidimensional approach is proposed that combines ethics with regulation, innovation, and education. It highlights the importance of developing detailed ethical frameworks, incorporating ethics into the training of professionals, conducting ethical impact audits, and encouraging the participation of stakeholders in the design of AI.

In addition, four fundamental pillars are presented for the ethical implementation of AI in organizations: 1) Integrated values, 2) Trust and transparency, 3) Empowering human growth, and 4) Identifying strategic factors. These pillars encompass aspects such as alignment with the company's ethical identity, governance and accountability, human-centered design, continuous training, and adaptability to technological and market changes.

The conclusion emphasizes that ethics must be the cornerstone of any organization's strategy that seeks to incorporate AI, establishing a solid framework that ensures that technology is developed and used in a way that respects and promotes human values.

\end{abstract}

 \newpage
\tableofcontents
\newpage

\section{Technocriticism and Key Actors in the Age of AI}
In recent years, a critical discourse has emerged that questions the impact of digital technologies and artificial intelligence (AI) on contemporary society. These analyses challenge the prevailing narrative that all technological innovation is inherently positive and beneficial. Instead, they highlight the potential risks of excessive technologization that could erode human autonomy and even have a negative impact on civilization.

This critical, yet valid, approach examines how digitalization and the rise of generative AI are radically reshaping our lives, the economic structure, and the cultural landscape. Concepts like "siliconization" have been coined to refer to the omnipresent imposition of a technological logic that permeates all spheres of human existence \cite{esadin2020}

Although it would be inappropriate to classify these analyses within purely Luddite movements - reminiscent of the anti-technology rebellion of the first half of the 19th century - or technophobic streams, it is more accurate to situate them within a technocriticism colored by concepts such as the end of Schumpeter's "creative destruction", which rejects the indiscriminate fusion of humans and technology under a skeptical gaze. This approach focuses mainly on analyzing the dominant trends and influential actors influencing the development and deployment of technology, with special attention to the role of the major technology corporations.

The perspective of these authors is mainly devoted to analyzing trends and actors as influential in a deterministic way, but this approach reveals one of the main limitations of such arguments: if accepted without further critical examination, we could wrongly conclude that key actors in the technology ecosystem play a passive or simply reactive role in this process.

The reality may be more complex and dynamic. Empirical studies show that corporations have been very decisive, not merely reactive, players in promoting certain models of technological innovation aligned with their business interests \cite{jobinai2019}. And while it is undeniable that large corporations do have considerable impact on the trajectory of AI development and application, it is crucial to recognize the equally pivotal role played by other actors such as governments, the scientific community, and organized civil society.

Let's talk about the probable influence or role that each one plays:
\begin{itemize}
  \item \textbf{Governments}: Through regulation and legislation, governments can and are trying to shape the development of AI. This includes everything from data privacy regulations to specific rules on the use of AI in sectors such as healthcare and transportation.
  \item \textbf{Scientists and Academics}: The academic world contributes not only through technological advancements but also through ethics in AI, social impact studies, and proposing alternatives to the technological status quo.
  \item \textbf{Civil Society}: Advocacy groups and individuals have the power to influence the development of technology through awareness campaigns, political pressure, and demands for ethical standards.
  \item \textbf{Technology companies}: Although many large companies may push a specific technological agenda, there are also numerous startups and companies seeking to develop and promote more ethical and human-centered AI approaches.
\end{itemize}

Therefore, while significant risks and trends can be pointed out, it is also essential to recognize and empower the active role that all actors in the AI ecosystem can and should have to ensure technological development that is aligned with human and social values. 

\section{How can organizations participate}
The approach of emphasizing the creation and adoption of \textit{digital values} to guide the development and use of artificial intelligence is important and must be proactive. From this position, it can be recognized that just as technology is not inevitable or destructive \textit{per se}, it is also not neutral and that the way it is developed, implemented, and used can—and should—be aligned with explicit ethical principles that reflect universal human values. 

\paragraph{KEY DIMENSIONS}\mbox{} \\

The main dimensions of how this vision can be implemented from the effective perspective of an organization can be:

\begin{itemize}
  \item \textbf{Development of ethical frameworks}: Academic institutions, regulatory bodies, and industry leaders can collaborate in creating detailed ethical frameworks for AI. These frameworks would include principles such as transparency, fairness, non-discrimination, and accountability. Examples of these efforts include the European Union's ethical guidelines for trustworthy AI. \cite{comisioneuropea2019}
  \item \textbf{Education and training in ethics for technologists}: Incorporating ethics into the education of software engineers, data scientists, project managers, and technical leaders is vital. This can help future professionals be aware of the implications of their work and equip them with tools to make technology decisions informed by ethical values.
  \item \textbf{Ethical impact audits and assessments}: Before implementing new AI systems, conducting ethical impact audits could be mandatory. This would help identify and mitigate potential negative effects on society or specific groups, ensuring that AI systems align with desired social values.
  \item \textbf{Integrationn of the user and stakeholder perspective}: Designing participatory processes where communities affected by AI technologies can provide their vision and concerns. This could help ensure that AI systems are inclusive and representative of the needs and values of a wider range of society.
  \item \textbf{Certifications and standards for ethical AI}: Developing standards and certifications that can demonstrate compliance with ethical norms in AI products and services. This would not only increase trust in these systems but could also motivate companies to consider ethics as an essential factor in AI design.
  \item \textbf{Multidisciplinary teams}: Some of the new challenges of these technologies, mainly generative models, have to do with the need to obtain a multi-professional appreciation: we not only need specialists in software, data, and science, but we will also require linguists, journalists, philosophers, psychologists, anthropologists, to name a few, and these will be according to the types, scopes, and dimensions of the projects in which they are involved. The very convergence of ethics with technologies that involve responses sometimes indistinguishable from human creativity, requires understandings and languages beyond the technological even.
\end{itemize}

\paragraph{COMPLEMENTARY DIMENSIONS}\mbox{} \\

The fact of ethics taken through value-driven design as a central focus is a must in the artificial intelligence debate, but there are other complementary ways that can also address the risks associated with AI, broadening the perspective and providing holistic solutions. These ways include:

\begin{itemize}
  \item \textbf{Regulation and legislation}: Beyond ethical principles, robust and specific regulation may be necessary to establish clear and enforceable limits on how AI can be developed and used. This could include legislation on data privacy, the use of facial recognition technologies, and limitations on automation in certain critical sectors.
\end{itemize}

\begin{itemize}
  \item \textbf{Human-centered design}: Integrating design approaches that prioritize human needs, capabilities, and limitations. This ensures that AI systems are developed in a way that complements human abilities and fosters effective collaboration between humans and machines, rather than replacing or displacing human work.
\end{itemize}

\begin{itemize}
  \item \textbf{Traceability and interpretability}: Developing and promoting technologies that are not only efficient but also transparent in their decision-making processes. Generative artificial intelligences allow receiving feedback on how results are reached and what values were influential, and this added value allows users to understand and trust how decisions are made by intelligent and autonomous systems and directly impacts the acceptance and capacity for effective supervision of these systems.
\end{itemize}

\begin{itemize}
  \item \textbf{Open and collaborative innovation}: Promote models of AI development that are inclusive and collaborative, involving multiple actors from different sectors and disciplines. Open innovation can help disseminate knowledge and best practices, and ensure that a wider range of interests and concerns are taken into account during technology development.
\end{itemize}

\begin{itemize}
  \item \textbf{Sustainable development}: Aligning the development of AI with the UN Sustainable Development Goals and other national, regional, or local ethical frameworks to ensure that it contributes positively to global and local challenges and, from here: to reducing inequality, improving health and well-being, and protecting the environment of the communities where the organization is inserted.
\end{itemize}

These avenues will never be mutually exclusive, and in fact, will be most effective when implemented in a coordinated manner. Addressing the challenges of AI requires a multifaceted approach that combines ethics, regulation, innovation, and education to create a technological ecosystem that is safe, fair, and beneficial to all.

Therefore, although we can discuss and apply multiple strategies to address the specific challenges presented by AI, these strategies are effectively facets of a broader commitment to values. Each of the approaches mentioned—from regulation to human-centered design and traceability—seeks to incorporate and reflect these ethical values into concrete practices.

\begin{figure} [H]
  \centering
  \includegraphics[width=1.3\linewidth]{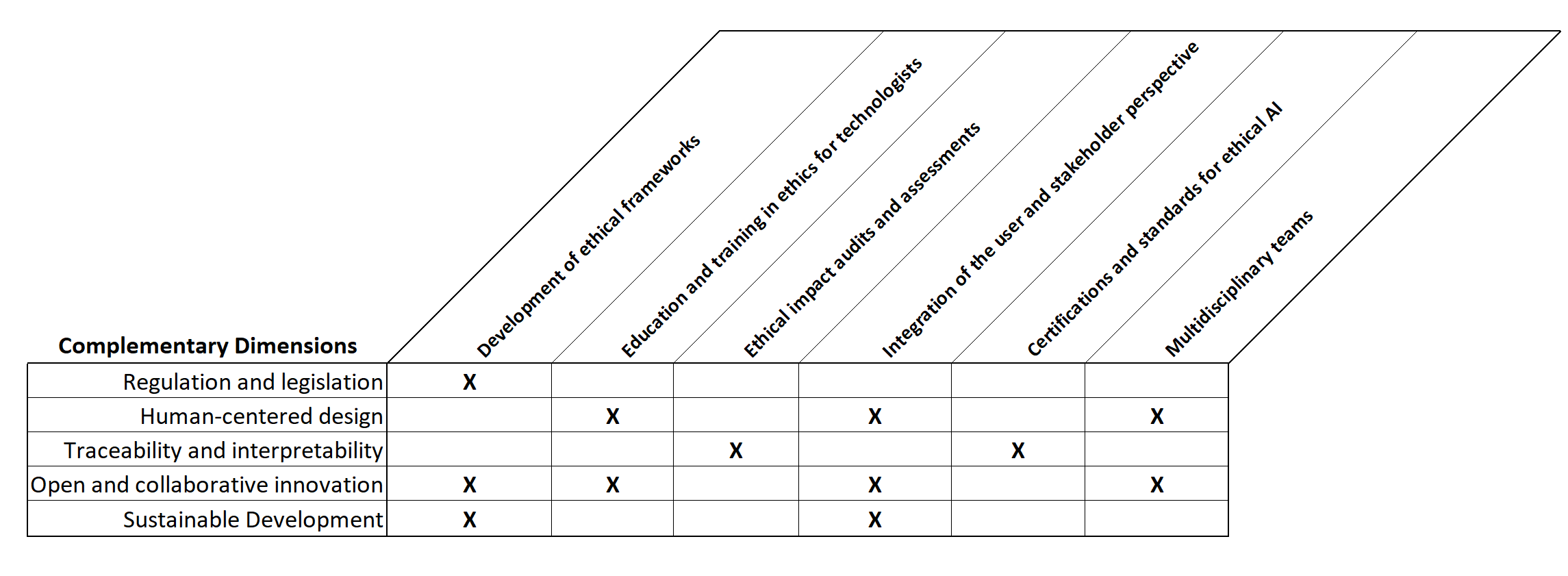}
  \caption{Potential intersections between key and complementary dimensions}
  \label{fig:claves-complementarias}
\end{figure}

This table (\textit{Figure 1}) illustrates the interaction between key and complementary dimensions. as additional considerations that reinforce an ethical implementation of AI. By using these connections, organizations can identify which complementary measures are essential to support each key dimension, facilitating a holistic approach to AI adoption.

\paragraph{SPECIFIC ACTIONS}\mbox{} \\

Specific actions are required that organizations can implement to integrate ethical principles into the use of Artificial Intelligence. These concrete measures ensure that the development and application of AI are aligned with fundamental ethical values such as fairness and transparency, thus strengthening accountability and trust in emerging technologies.

\paragraph{\textbf{Action: Define ethical principles for AI}}
\begin{itemize}
  \item Form an ethics committee that involves leaders, experts, and diverse stakeholders to define the core ethical principles that will guide the use of AI in the organization.\cite{trejo2020hacia}
  \item Develop and formally document an AI ethics code that establishes principles such as fairness, transparency, accountability, and respect for privacy as mandatory guidelines.
  \item Integrate this ethics code into the company's mission, values, and strategic objectives, ensuring that it is widely communicated and understood by all levels.
\end{itemize}

\paragraph{\textbf{Action: Implement ethical AI design practices}}
\begin{itemize}
  \item Adopt agile and human-centered methodologies that involve employees, customers, communities, and diverse stakeholders from the early stages of AI solution design.
  \item Incorporate ethical and social impact assessments as an integral part of the design process, identifying and mitigating potential risks before development and implementation.
  \item Form multidisciplinary teams that include technical, business, ethics, privacy, and representatives of impacted groups for a comprehensive approach.
\end{itemize}  

\paragraph{\textbf{Action: Ensure responsible AI implementation}}
\begin{itemize}
  \item Establish a governance framework, policies, and procedures to continuously monitor compliance with ethical principles throughout the AI systems life cycle.
  \item Conduct regular ethical audits by independent entities to assess algorithms, data, processes, and outcomes for bias, risks, or non-compliance.
  \item Implement technical traceability and interpretation solutions for processes that allow understanding the reasons behind results and decisions that directly affect people.
\end{itemize}

\paragraph{\textbf{Action: Conduct continuous evaluations and adjustments}}  
\begin{itemize}   
  \item Create accessible channels (e.g., ethics lines, online communities) for employees, customers, and the public to report concerns, observed biases, or other ethical impacts.
  \item Form review committees that analyze these reports, investigate thoroughly, and recommend corrective actions including adjustments to practices, processes, or AI systems.
  \item Schedule regular external social and ethical impact evaluations that allow identifying improvement opportunities and trends that require strategic changes.
\end{itemize}

\paragraph{\textbf{Action: Implement education and training programs}}  
\begin{itemize}
  \item Develop mandatory continuous training plans at all levels, combining technical knowledge of AI with modules on ethics, bias, privacy, and human rights.
  \item Incorporate practical activities, case studies, and ethical dilemmas that allow employees to apply the principles learned to real-world situations.
  \item Establish internal communication channels, events, and campaigns that continuously reinforce the importance of ethics and the responsible use of AI for the common good.
\end{itemize}

\begin{figure}[H]
  \centering
  \includegraphics[height=0.65\linewidth]{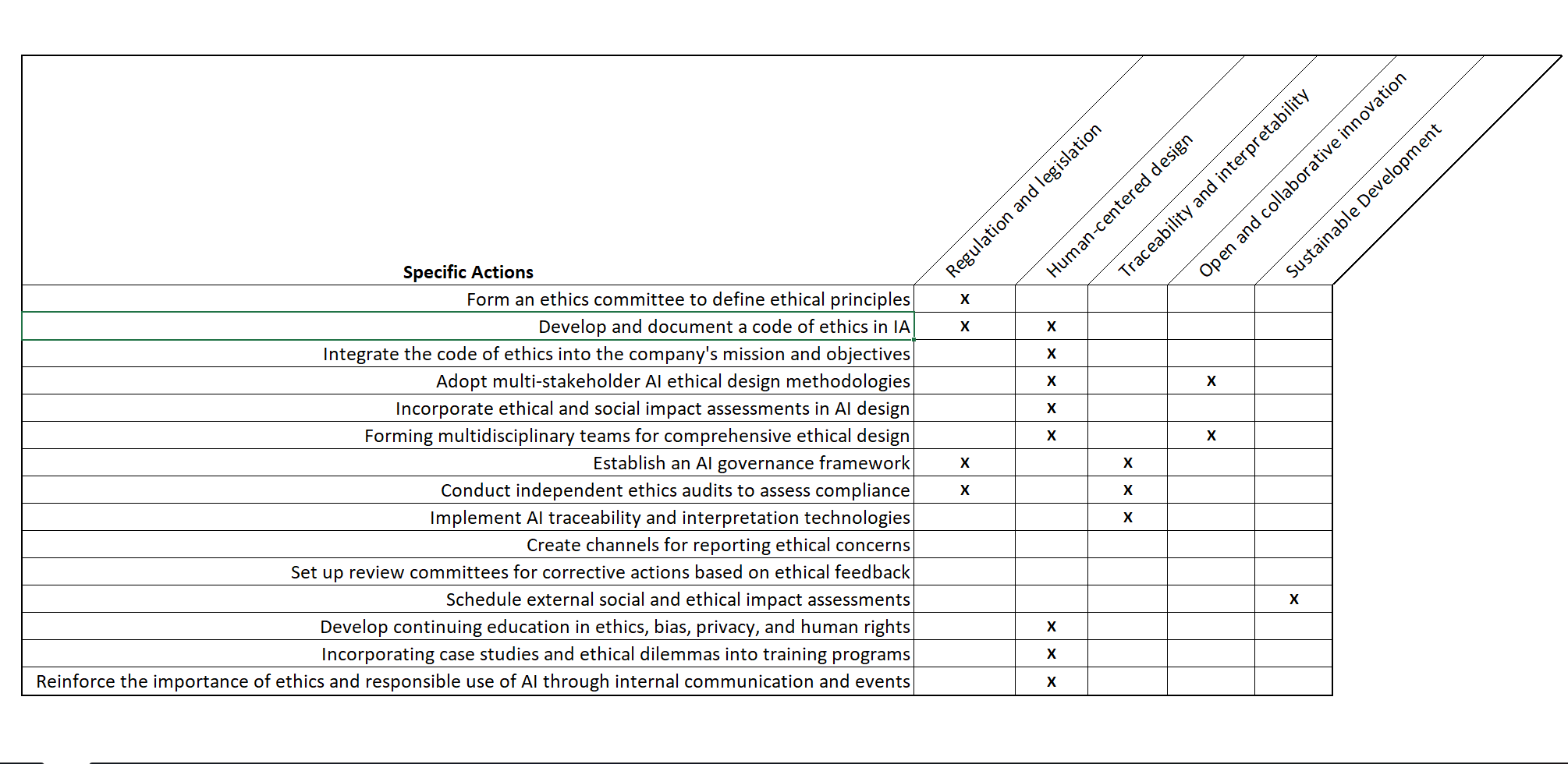}
  \caption{Potential actions corresponding to the key dimensions within an AI adoption strategy}
  \label{fig:dimensiones-acciones}
\end{figure}

The table in Figure 2 is a proposal of what specific actions can be applied to strengthen the key dimensions in the implementation of AI. It shows how concrete interventions, such as defining ethical principles or implementing ethical design practices, align with fundamental areas such as the development of ethical frameworks, education in ethics for technologists, and ethical impact audits.

\paragraph{INCLUSION AND DIVERSITY}\mbox{} \\

As a separate and distinct element of the general methodological context, we explain this "Gender and Diversity" section to highlight the transversal and critical relevance of these aspects in the implementation of artificial intelligence. These issues are not simply additional or secondary; they are fundamental and must be integrated into all dimensions to ensure equity, effectiveness, and justice in the adoption or development of AI. 

 As AI becomes a more integral tool within an organization and its processes, the need to design systems that reflect and respect internal diversity in particular, but the diversity of society in general, becomes more critical than ever. 
 
 One very well-known component that can be used to positively regulate is the training algorithms of AI models, which are often developed and trained using datasets that do not take into account gender, ethnicity, and other sociodemographic factors, which can result in unintentional biases that perpetuate discrimination.
 
 For example, facial recognition systems have shown significantly higher error rates for women and people of color. These technological biases are not just technical errors; they reflect and perpetuate existing inequalities in society. Let's look at some known and not-so-obvious biases:
\begin{itemize}
  \item \textbf{Racial and ethnic bias}: Algorithms that have not been properly adjusted may show preferences for certain racial or ethnic groups, due to unequal representation in training data. For example: Online dating apps may be prone to racial bias and reinforce stereotypes. Some of these apps have been criticized for racial discrimination present in the choice of partners.
  \item \textbf{Age bias}: Discrimination towards individuals of certain ages, particularly in hiring and marketing decisions, where AI could favor users of certain age ranges, ignoring the needs and preferences of other groups.
  \item \textbf{Social and economic class bias}: Automated decisions that favor individuals of certain socioeconomic strata. Credit assessment algorithms can perpetuate financial exclusion of certain groups, preventing access to loans and economic opportunities. For example, it was discovered that a major credit card company assigned lower credit limits to women, even when their financial profiles were similar to those of men.
  \item \textbf{Linguistic and cultural bias}: Algorithms that do not adequately recognize or value linguistic and cultural diversity, leading to inadequate or irrelevant responses for users of different cultures or who speak different languages, especially minorities.
  \item \textbf{Geographical bias}: AI could show preferences for users of certain geographical locations, based on the prevalence of data from those places in the training set, thus neglecting users in less represented regions.
\end{itemize}

\paragraph{Strategies to promote equity}\mbox{}\\
To combat these challenges, organizations must adopt specific strategies that ensure inclusion and equity in the development and use of AI \cite{unesco2023}. Some of these strategies are:

\begin{itemize}
  \item \textbf{Diversification of Datasets:} Ensure that the data used to train AI algorithms is representative of global demographic diversity. This includes collecting data that is sufficiently varied and conducting rigorous tests to identify and correct biases before algorithms are deployed.
   
  \item \textbf{Inclusive Development Teams:} Foster diversity within AI development teams. Diverse teams not only enhance innovation and creativity but can also potentially identify bias and equity issues that could go unnoticed in more homogeneous teams.
   
  \item \textbf{Regular Equity Audits:} Implement periodic reviews of AI systems to assess their performance and fairness. Audits should be conducted both internally and by independent third parties to ensure transparency and objectivity.
   
  \item \textbf{Education and Awareness:} Educate all employees about the importance of diversity and inclusion in technology. This includes specific training on how biases can infiltrate algorithms and the ways to mitigate these risks in the design and implementation of AI systems.
   
  \item \textbf{Community Participation:} Involve diverse communities in the AI development process. Obtaining feedback from a wide range of users can help identify equity issues and improve the acceptance and effectiveness of AI systems.
   
\end{itemize}
\section{Four Pillars for Implementing an Ethical Framework in Organizations}

While understanding the various dimensions, actors, or actions presented is fundamental to holistically addressing the ethical challenges of artificial intelligence, it is equally essential to have a structured methodological approach. This framework will allow organizations to coherently integrate ethical principles into each stage of the AI systems lifecycle, from strategy to daily operations.

In this line, four fundamental pillars are proposed that will act as guiding principles for any organization to build a robust and comprehensive approach oriented towards the ethical incorporation of Artificial Intelligence. These pillars provide a methodical and unified vision, bringing together the diverse ethical, social, and governance components previously analyzed into a practical and actionable framework for organizations.

\begin{figure}[H]
  \centering
  \includegraphics[width=1\linewidth]{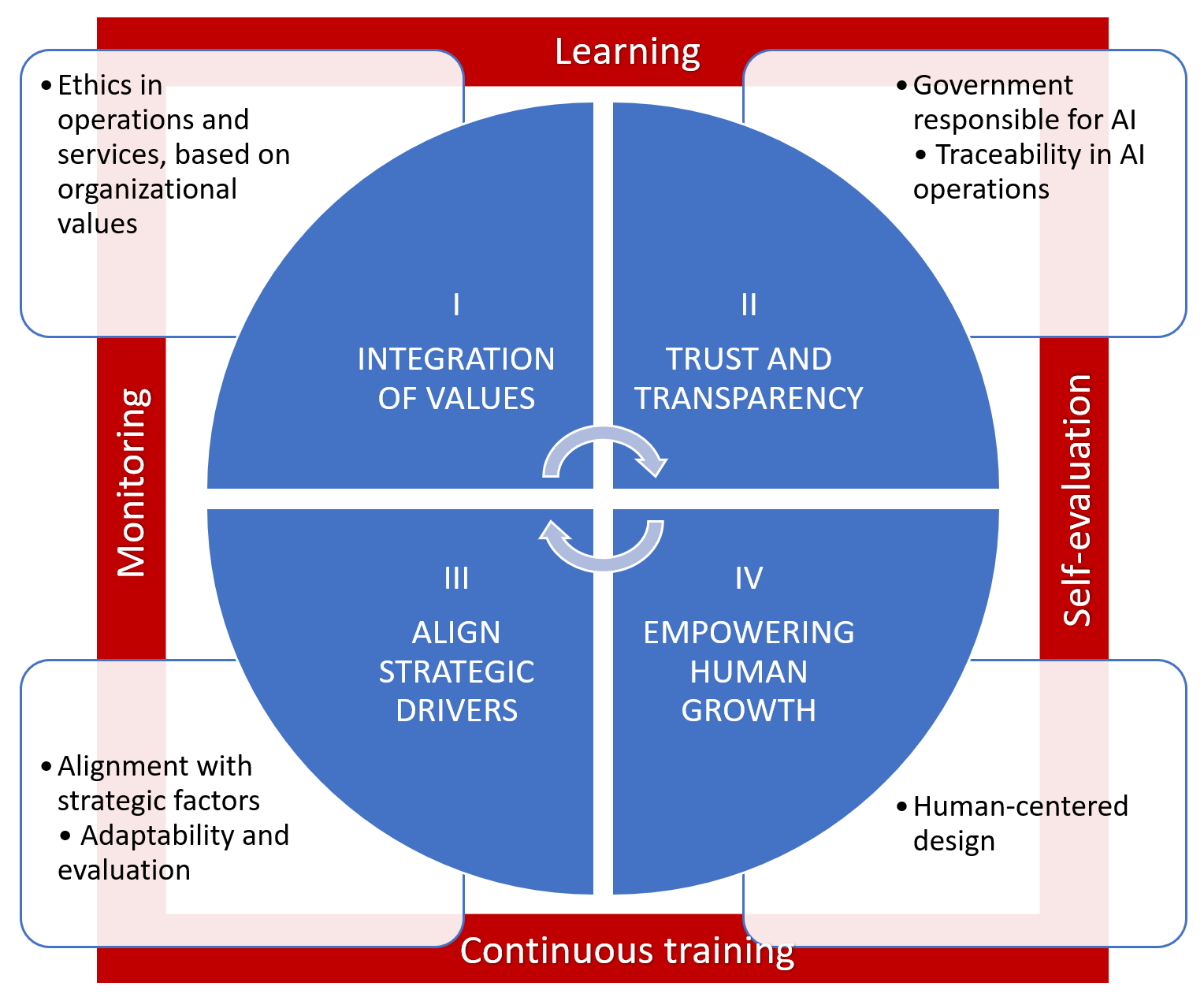}
  \caption{\textit{Pillars for an ethical framework for the adoption and development of Artificial Intelligence in organizations}}
\end{figure}

\paragraph{PILLAR I: INTEGRATION OF VALUES:}\mbox{}\\

This pillar focuses on embedding the company's values into all aspects of AI development and use, ensuring that technologies reflect and promote the organization's identity and ethical standards. Continuous training is a critical requirement to keep all employees aligned with these values. It is not simply a matter of establishing ethical principles in documents but of making these values intrinsic to the organizational culture and reflected in every stage of AI development and use.

The key lies in involving all employees, from leadership to technical and operational teams, in a process of constant review and alignment with the company's ethical identity. This involves continuous training programs that not only convey technical knowledge but also a deep understanding of how to apply organizational values in decision-making and the design of AI solutions.

Additionally, it is essential to establish processes and policies that translate these values into concrete practices. For example, audits of algorithms to detect and mitigate bias, data privacy protocols, or accountability mechanisms for potential negative impacts. In this way, values cease to be mere declarations and become operational guidelines that guide the daily work of all those involved in the development of AI.

\paragraph{PILLAR II: TRUST AND TRANSPARENCY:}\mbox{}\\

Emphasizes AI governance, transparency, and accountability. "AI ambassadors" are key to monitoring and ensuring the quality and diversity of data, as well as providing continuous feedback, thus improving the reliability and equity of AI systems.

Solid governance structures must be implemented, which include ethics committees, independent audits, and continuous monitoring mechanisms. These bodies must ensure compliance with ethical principles and guarantee that AI systems operate fairly, transparently, and responsibly.

Another essential aspect is the development of interpretable and traceable AI systems, which allow users to understand the reasons behind the decisions made and the factors that influence the results. This not only promotes trust in AI systems but also facilitates the identification and correction of potential biases or errors by being able to trace the data, rules, and processes that led to certain results. 

Finally, it is crucial to establish open and two-way communication channels with all stakeholders, including employees, customers, and affected communities. This allows for valuable feedback to be received, concerns to be addressed, and a constant dialogue to be maintained about the impact of AI on society.

\paragraph{PILLAR III: EMPOWERING HUMAN GROWTH:}\mbox{}\\

By focusing on the collaboration between human and artificial intelligence, the implementation of this pillar seeks to enhance creativity and minimize monotonous tasks, highlighting the importance of adapting to new roles and continuous learning. The goal is to leverage the unique strengths of AI to enhance human capabilities, freeing people from repetitive tasks and allowing them to focus on more creative and rewarding activities.

Human-centered design is a key decision that involves the active participation of users and employees in the development of AI solutions. This ensures that systems adapt to human needs and limitations, rather than forcing people to adapt to technology.

Additionally, it is essential to invest in continuous training and development programs to prepare the workforce for the changes driven by AI. This includes the development of complementary skills, such as critical thinking, problem-solving, and effective communication, which will be essential in an increasingly automated work environment.

\paragraph{PILLAR IV: ALIGN STRATEGIC DRIVERS:}\mbox{}\\

While ethics and values must be the foundation of any AI initiative, this pillar recognizes the importance of aligning these technologies with the organization's strategic objectives. It is about identifying and leveraging key factors that can enhance the use of AI to drive growth, efficiency, and competitiveness for the company.

This requires an adaptable and practical approach that involves constantly monitoring technological and market trends. Organizations must be prepared to adjust their strategies and adopt new AI solutions as opportunities or challenges arise.

And it is essential to conduct comprehensive assessments of the potential risks and benefits of each AI implementation. This involves considering not only the financial impacts but also the ethical, social, and environmental aspects, ensuring that innovation is carried out responsibly and sustainably.

\section{Conclusions}
The ethical implementation of artificial intelligence in organizations is more than a responsibility; it is an essential strategy that must precede even the definition of business objectives. Establishing rigorous ethical frameworks not only guides the development and use of AI but also protects the organization against future risks and strengthens its long-term reputation. These frameworks are not simply a complement to technological and commercial strategies, but the foundations upon which they are built.

For today's organizations, the task of integrating ethics into AI involves several key actions. First, it is important for senior management to adopt and promote an ethical vision that is integrated into all phases of the AI lifecycle, from conceptualization to deployment and operational review. This means that AI decisions must be reviewed not only for their economic impact but also for their alignment with ethical principles such as transparency, fairness, and respect for privacy and human dignity.

Secondly, organizations must be proactive in training their teams, not only in technical skills but also in ethical competencies. This includes understanding the social implications of AI, the ability to identify potential biases in algorithms and data, and the development of solutions that respect the rights and values of all stakeholders. Ethics training should be a continuous part of corporate education, adapting to the new trends and challenges that emerge as technology evolves.

Additionally, the implementation of continuous auditing and evaluation processes is fundamental to ensure that AI systems operate within the established ethical boundaries. These processes must include not only internal reviews but also the participation of independent third parties who can offer fresh perspectives and help avoid "organizational blindness" to ethical problems.

Finally, transparency with consumers and society in general is essential. Organizations must openly and honestly communicate how they use AI, the data they collect, and the controls they have in place to protect that data and ensure fair and equitable treatment. This is not only an ethical practice but also contributes to building trust, an invaluable asset in the digital age.

Overall, ethics must be the cornerstone of any organization's strategy that aspires to incorporate new artificial intelligence technologies. Before pursuing the business potential that AI can offer, it is imperative to establish a solid framework that ensures that technology is developed and used in a way that respects and promotes human values. By doing so, organizations not only secure their own sustainable future but also contribute to the well-being of society as a whole.

\newpage

\printbibliography[
heading=bibintoc,
title={Bibliography}
]

\end{document}